# Enhancing Power System Cyber-Security with Systematic Two-Stage Detection Strategy

Xingpeng Li, *Member, IEEE* and Kory W. Hedman, *Senior Member, IEEE*

*Abstract*— State estimation estimates the system condition in real-time and provides a base case for other energy management system (EMS) applications including real-time contingency analysis and security-constrained economic dispatch. Recent work in the literature shows malicious cyber-attack can inject false measurements that bypass traditional bad data detection in state estimation and cause actual overloads. Thus, it is very important to detect such cyber-attack. In this paper, multiple metrics are proposed to monitor abnormal load deviations and suspicious branch flow changes. A systematic two-stage approach is proposed to detect false data injection (FDI) cyber-attack. The first stage determines whether the system is under attack while the second stage identifies the target branch. Numerical simulations verify that FDI can cause severe system violations and demonstrate the effectiveness of the proposed two-stage FDI detection (FDID) method. It is concluded that the proposed FDID approach can efficiently detect FDI cyber-attack and identify the target branch, which will substantially improve operators' situation awareness in real-time.

## NOMENCLATURE

**Sets**
$K$    Set of branches.
$KA$    Set of branches that have the top ten values for malicious load deviation index.
$K(n-)$    Set of branches with bus $n$ as from-bus.
$K(n+)$    Set of branches with bus $n$ as to-bus.
$N$    Set of buses.
$NL$    Set of load buses.
$NL(k)$    Set of load buses that are critical to branch $k$.

**Indices**
$k$    branch.
$n$    bus.
$n(k-)$    From-bus of branch $k$.
$n(k+)$    To-bus of branch $k$.

**Parameters**
$d_{n0}$    Actual load at bus $n$ at $t = 0$.
$d_{n0,M}$    Load measurement at bus $n$ at $t = 0$.
$d_{n-}$    Actual load at bus $n$ at $t = -\Delta T$.
$l$    Target branch $l$ of FDI attack.
$L_S$    Load shift factor.
$Limit_k$    Thermal limit of branch $k$.

$N_1$    Limit of an $l_1$-norm constraint.
$NL_k$    Number of load buses that are critical to branch $k$.
$P_{l,0}$    Actual flow on target branch $l$ at $t = 0$.
$P_{k0}$    Actual flow on branch $k$ at $t = 0$.
$P_{k0,M}$    Measurement of flow on branch $k$ at $t = 0$.
$P_{k-}$    Actual flow on branch $k$ at $t = -\Delta T$.
$P_{k+,SCED}$    Scheduled flow on branch $k$ at $t = \Delta T$, determined by SCED that runs at $t = 0$.
$PTDF_{n,k}$    Power transfer distribution factor for branch $k$ due to an injection change at bus $n$.
$\Delta T$    Interval of a period.
$x_k$    Reactance of branch $k$.
$\Delta d_n$    Actual load difference at bus $n$ between $t=0$ and $t=-\Delta T$.

**Variables**
$c$    Attack vector of bus phase angles.
$\Delta \tilde{d}_n$    Malicious load deviation at bus $n$.
$p_l$    Post-attack actual flow on target branch $l$.
$\tilde{p}_l$    Post-attack cyber flow on target branch $l$.
$p_k$    Post-attack actual flow on branch $k$.
$\tilde{p}_k$    Post-attack cyber flow on branch $k$.
$\Delta p_l$    Difference between the post-attack actual power flow and cyber power flow on the target branch $l$.
$\Delta p_k$    Difference between the post-attack actual power flow and cyber power flow on branch $k$.
$\theta_n$    Post-attack actual phase angle of bus $n$.
$\tilde{\theta}_n$    Post-attack cyber phase angle of bus $n$.

**Functions**
$sgn(x)$    1 if $x$ is positive; 0 if $x$ is zero; -1 if $x$ is negative.
$size(S)$    Number of elements in the set $S$.

## I. INTRODUCTION

In modern power systems, energy management systems (EMSs) are used to help system operators manage real-time operations. Key functions of EMS include state estimation (SE), real-time contingency analysis (RTCA) and real-time security-constrained economic dispatch (SCED). These functions execute in a coordinated way such that the system reliability can be maintained with least-cost solutions.

State estimation executes routinely in real-time and serves as a core function in EMS for monitoring system condition. For instance, the SE at PJM, an independent system operator (ISO) in the United States, runs on a one-minute basis and can converge in 30-45 seconds [1]. With the measurement data received from remote terminal units (RTU) or local control centers through a communication network, state estimation can effectively estimate the system status and provide a basis for other subsequent applications in real-time. State estimation methods are based on the physical relationships of the power system and require data redundancy.

The traditional bad data detection and identification module of SE can detect random bad data that are introduced by large

The research presented in this manuscript is funded by the National Science Foundation (NSF) Award (1449080).

Xingpeng Li is with the Department of Electrical and Computer Engineering, University of Houston, Houston, TX 77204 USA (e-mail: xingpeng.li@asu.edu). Kory Hedman is with the School of Electrical Computer, and Energy Engineering, Arizona State University, Tempe, AZ 85287 USA (email: kwh@myuw.net).

measurement errors. It ensures the impact of random measurement noises on SE is minimal. However, recent work [2]-[8] in the literature shows that malicious cyber-attack can inject false measurements that are designed to meet the physical laws and bypass bad data detection. This indicates that power system state estimation is subject to false data injection (FDI) cyber-attack that may lead to physical consequences. Moreover, real-world cyber-attack example exists: in 2015, a cyber-attack against the Ukraine power grid caused a serious physical blackout event, which is referred to as the first known successful power system incident caused by a cyber-attack [9].

FDI cyber-attack on power system state estimation has gained significant attention since it was first proposed by Liu *et al*. in [2]: FDI cyber-attack injects coordinated false measurements in an unobservable way such that the SE solutions are corrupted and may negatively affect operators real-time dispatch decisions. The attack model proposed in [2] and its impact are further analyzed, and more detailed results are presented in [3], as well as a generalized FDI cyber-attack model. Two regimes of attacks, a strong regime and a weak regime, are presented in [4]. The strong regime attack with access to a sufficient number of meters can launch unobservable attacks while the weak regime attack can be detected as only a limited number of meters are under control of the attacker. A graph theory based algorithm is proposed in [5] to identify the locations where attackers can attack with the least-number measurements to keep the attack from being detected by AC state estimation. Thus, those locations may need more protection against potential FDI attacks. It is shown in [6] that the attacker can launch an unobservable attack by only introducing false measurements within a subgraph that is determined by the subgraph algorithm proposed in [5]. Extended on [6], a bi-level optimization is proposed in [7] to maximize the physical flow on a target branch, which is equivalent to maximizing the branch overload. Though the FDI cyber-attack approach proposed in [7] can cause unobservable branch overloads, it does not scale due to computational complexity. Therefore, three computationally efficient algorithms are proposed in [8] to speed-up the solution time and provide boundaries on system vulnerability.

As illustrated in [2]-[8], power systems are subject to FDI cyber-attack and the attacker can invisibly compromise SE. Thus, SE under FDI cyber-attack may be corrupted and provide a biased base case for other EMS applications, which poses substantial risks on RTCA and SCED. A biased system condition may mislead system operators to take incorrect actions such as improper generation adjustment, which leads to severe violations or damage to the power system. Therefore, it is critical to ensure the SE results are correct. Developing FDI detection (FDID) strategies that can efficiently detect FDI cyber-attack is vital for reliability enhancements and secure operations of electric power systems. A number of methods have been proposed in the literature to address system vulnerabilities due to FDI cyber-attack. These methods can be grouped into two categories [10]-[11]: (i) protection-based methods [12]-[19] and (ii) detection-based methods [20]-[25].

A greedy algorithm is proposed in [12] to select measurements considering a budget constraint such that those measurements will be encrypted for protection to maximize the system security. Reference [13] designed a fast greedy algorithm to protect a subset of measurements for defending FDI cyber-attack. Known-secure PMUs are used to defense against FDI cyber-attack in [14]-[15], where a scheme is proposed to find the minimum number of necessary PMUs. A least-budget defense strategy, which can achieve quality solutions in a reasonable time, is proposed in [16] to enhance state estimation against FDI cyber-attack by protecting critical meters. In [17], a specific set of measurements are selected and protected to detect FDI cyber-attack. The smallest set of measurements can be determined by the proposed two approaches, brute-force search and protecting basic measurements. A novel countermeasure against FDI cyber-attack is proposed in [18] by protecting critical state variables, which is realized by protecting a minimum number of meter measurements; further, graphical methods are used to determine the minimum number of necessary measurements [19]. The work presented in [12]-[19] are protection-based methods that share two drawbacks [11]: (i) data redundancy decreases; and (ii) protection is not absolutely secure. SE can work more effectively with higher degree of redundancy; with a subset of trusted measurements only, the robustness and accuracy of SE solutions will drop. Moreover, the attacker may be able to compromise the protection that leads to the failure of protection-based methods.

A novel FDID mechanism that uses nuclear norm minimization and low rank matrix factorization is proposed in [20] to separate nominal grid states and the anomalies. The Bayesian framework proposed in [21]-[22] can preserve and trace possible system states using prior information, which can detect false data by identifying statistically unlikely measurements. It is shown in [23] that random bad data injection can be identified by state estimation while stealth bad data injection can bypass state estimation; thus, [23] proposes a defense strategy against the stealth bad data injection by conducting real-time statistical analysis on a sequence of data at the minimum cost of delay. A three-phase mechanism is proposed in [24] to detect FDI cyber-attack by evaluating spatiotemporal correlation between system states. A centralized FDI detector that is based on the generalized likelihood ratio and a distributed FDI detector that employs the adaptive level-triggered sampling technique are proposed in [25] to efficiently detect FDI cyber-attack. The above methods [20]-[25] are classified as detection-based methods that do not rely on the protection on pre-selected key measurements. However, those methods may not be able to detect the cyber-attack when the injected false data fit the distribution of previous measurements.

In this paper, we propose a real-time systematic two-stage detection approach against FDI cyber-attack, which relies on neither key meter protection nor historical data. The proposed FDI detection method is based on an alert system with multiple novel metrics; sensitivity factors such as power transfer distribution factors (PTDF) are used to calculate those devised metrics. With the proposed strategy, majority of the cyber-attack events can be successfully detected in real-time. The contributions of this work are presented as follows:

(i) The proposed two-stage FDID strategy is computationally tractable. The first stage determines whether the system is under FDI cyber-attack and the second stage identifies the target branch.
(ii) The proposed metrics and alert system can effectively monitor abnormal load deviations and flow changes.

(iii) The proposed approach can enhance SE by effectively detecting FDI cyber-attack in real-time and, thus improve system cyber-security. The proposed systematic FDID scheme is more robust since it depends on neither meter protection nor historical data.

(iv) Simulation results demonstrate the proposed FDID scheme can efficiently detect FDI cyber-attack and identify the target branch; the results also show the false alarm rate and false dismissal rate are low.

The rest of this paper is organized as follows. Section II briefly introduces state estimation. Section III explains FDI cyber-attack. Section IV presents the proposed monitoring metrics, FDI cyber-attack alert system and systematic two-stage FDID approach. Section V presents the numerical results for FDI and FDID. Finally, Section VI concludes the paper and Section VII presents the future work.

## II. STATE ESTIMATION

State estimation processes the data received from RTUs or local control centers and estimates the system status in real-time. It provides a base case or a starting point for other EMS functions such as RTCA and SCED. Thus, state estimation is an essential function of EMS as it is the basis for other modules in EMS. The sequence of power system real-time operations is illustrated in Fig. 1.

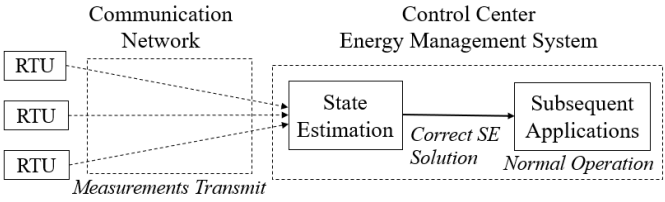
Fig. 1. Power system real-time operation sequence.

SE is run continuously in real-time to estimate the system status including bus voltages and branch flows. The measurement model for SE can be represented by (1). In (1), $e$ denotes measurement error vector and $h(x)$ describes the relationship between state variables $x$ and measurements $z$.

$$z = h(x) + e \quad (1)$$

In this work, the DC model is used to illustrate the proposed work. For DC state estimation, the relationship between state variables $x$ and measurements $z$ is linear and then, $h(x)$ can be replaced by $Hx$, where $H$ is a constant measurement function coefficient matrix. Thus, (1) can be replaced by (2) for DC state estimation. In (2), state variable $x$ denotes nodal phase angle. This paper focuses on the DC model.

$$z = Hx + e \quad (2)$$

## III. FDI CYBER-ATTACK

As discussed in the previous section, it is critical to ensure SE functions as other EMS modules depend on SE solutions. However, prior work in literature shows that FDI cyber-attack can compromise SE and cause unobservable branch overloads. Fig. 2. illustrates how a cyber-attack compromises the system by replacing true measurements with false measurements.

To launch an unobservable FDI cyber-attack, the injected false measurements should meet (3) that represents the measurement model under attack. In (3), $\tilde{x}$ denotes the state variable under attack and $\tilde{z}$ denotes the measurements under attack. Equation (4) defines the relationship between the actual state variable without attack and the cyber state variable under attack; variable $c$ is referred to as attack vector in this paper. Note that the attacker cannot directly modify the state variable; rather, the attacker injects false measurements that indirectly leads to biased estimates of the state variable. In addition, if the attacker targets at a specific state variable, then, all the measurements related to that state variable must be changed accordingly in order to cover such an attack.

$$\tilde{z} = H\tilde{x} + e \quad (3)$$
$$\tilde{x} = x + c \quad (4)$$

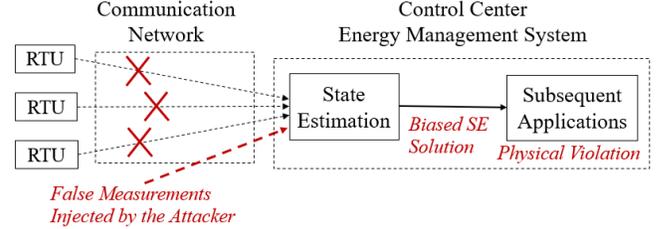
Fig. 2. Power system real-time operation under FDI cyber-attack.

Fig. 3 shows the time line for illustrating the FDI cyber-attack. There are two dispatch intervals shown in Fig. 3. In this work, it is assumed the attacker launches the FDI cyber-attack at $t = 0^-$ that is before the start of the second period and the system information at $t = -\Delta T$ is accurate. Due to operator's generation re-dispatch, the actual power flows will change in the second period. It is assumed that the attacker does not have the capability of manipulating the generation measurements; rather, the attacker can inject false load measurements, which is more realistic. Moreover, the attacker is assumed to have access to all load measurements, which represents the worst FDI cyber-attack scenario.

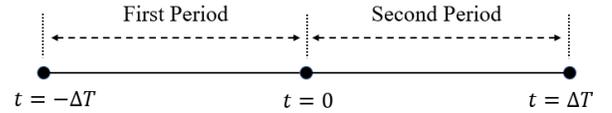
Fig. 3. Time line for illustrating the FDI cyber-attack.

In [7], a bi-level optimization model is proposed to determine the attack vector and false load vector that can cause the most severe loading level on a target transmission and may result in physical flow violation. To solve the problem in a timely manner, accelerating methods for providing fast solutions to this bi-level model are proposed in [8]. A heuristic method proposed in [8] is enhanced in this work by considering the line flow directions and the associated model with B-$\theta$ power flow formulation is presented below,

$$\text{maximize } sgn(P_{l,0})(p_l - \tilde{p}_l) \quad (5)$$

subject to

$$p_k = (\theta_{n(k-)} - \theta_{n(k+)})/x_k, \ k \in K \quad (6)$$
$$\tilde{p}_k = (\tilde{\theta}_{n(k-)} - \tilde{\theta}_{n(k+)})/x_k, \ k \in K \quad (7)$$
$$\tilde{\theta}_n = \theta_n + c_n, \ n \in N \quad (8)$$
$$\Delta \tilde{d}_n = \sum_{k \in K(n-)}(p_k - \tilde{p}_k) - \sum_{k \in K(n+)}(p_k - \tilde{p}_k), \ n \in NL \quad (9)$$
$$-L_S d_{n0} \leq \Delta \tilde{d}_n \leq L_S d_{n0}, \ n \in NL \quad (10)$$
$$-c_n \leq s_n, \ n \in N \quad (11)$$
$$c_n \leq s_n, \ n \in N \quad (12)$$
$$\sum_n s_n \leq N_1 \quad (13)$$

The objective of this model is to maximize the difference between post-attack physical and cyber power flows on a pre-specified target branch $l$. The attack problems for different target branches are independent and the case that leads to the most severe overload will be considered as the worst case. The term $sgn(x)$ in (5) denotes the sign function. Equations (6) and (7) calculate the post-attack physical and cyber branch flows respectively. Equation (8) shows the relationship between physical bus angles and cyber bus angles. Equation (9) calculates the malicious load deviation for each load bus while (10) ensures that the load shift is within limits. The summation of the absolute change in state variables is restricted by (11)-(13), which is equivalent to an $l_1$-norm constraint [7].

The above model can be further simplified by introducing a new variable $\Delta p_k$ that denotes the difference between post-attack physical and cyber power flows. Then, the simplified model can be represented by (10)-(17). This FDI cyber-attack model is implemented to provide the data required for the FDID studies in this paper.

$$maximize\ sgn(P_{l,0})\Delta p_l \quad (14)$$

subject to (10)-(13) and

$$\Delta p_k = (-c_{n(k-)} + c_{n(k+)})/x_k,\ k \in K \quad (15)$$
$$\Delta \tilde{d}_n = \sum_{k \in K(n-)}(\Delta p_k) - \sum_{k \in K(n+)}(\Delta p_k),\ n \in NL \quad (16)$$
$$\Delta p_k = p_k - \tilde{p}_k,\ k \in K \quad (17)$$

IV. FDID METHODOLOGY AND ALGORITHM

*A. FDID Metrics*

Two categories of metrics are proposed in this paper to effectively detect potential FDI cyber-attacks. They are the branch overload risk index (BORI) and the malicious load deviation index (MLDI). BORI monitors suspicious changes in branch flows and identifies potential overloads, while MLDI can recognize load change patterns and identify malicious load deviation. The proposed metrics BORI and MLDI are individual metrics for determining whether a specific branch is the attack target. Based on BORI and MLDI, systematic metrics and methodology are proposed, and they are presented in Section IV.B. In addition, this paper proposes an FDI cyber-attack alert system that has four different alert levels defined as *Danger*, *Warning*, *Monitor*, and *Normal*.

**Branch Overload Risk Index**

To execute an unobservable FDI cyber-attack that would overload a branch, the attacker can change the measurements including load measurements that are sent to the system operator. In the cyber world, the attacker can deliberately reduce the flow on a congested line or a heavily loaded line by shifting loads. This would mislead operators to believe that there is extra available capacity on the target branch; then, operators may re-dispatch generation to take advantage of that extra available capacity and reduce the total cost. However, in the real world, there is no such extra available capacity and physical overloads may occur. Thus, based on this type of flow change pattern, a branch overload risk index is proposed in this work to detect FDI cyber-attacks.

Since attackers may or may not consider the effects of generation re-dispatch, two similar but different metrics are proposed in this paper: $BORI1_k$ and $BORI2_k$ that are defined in (18) and (19) respectively. $BORI1_k$ only considers the flow changes in the previous interval while $BORI2_k$ takes SCED into account. A comprehensive metric $BORI_k$ is proposed to combine these two metrics. As shown in (20), $BORI_k$ is defined to be the larger value between $BORI1_k$ and $BORI2_k$.

$$BORI1_k = sgn(P_{k-})(P_{k-} - P_{k0,M} + P_{k-})/Limit_k \quad (18)$$
$$BORI2_k = sgn(P_{k-})(P_{k-} - P_{k0,M} + P_{k+,SCED})/Limit_k \quad (19)$$
$$BORI_k = max(BORI1_k, BORI2_k) \quad (20)$$

The alert level criteria for $BORI_k$ is defined in Table I. In this table, $ALB_k$ denotes the alert level associated with $BORI_k$ and it enables operators to determine whether a branch is under attack from the viewpoint of flow violations.

TABLE I ALERT LEVEL CRITERIA BASED ON $BORI_k$

| Alert level $ALB_k$ | $BORI_k$ |
|---|---|
| Danger | >115% |
| Warning | >110% |
| Monitor | >105% |
| Normal | <105% |

**Malicious Load Deviation Index**

Power transfer distribution factors are widely used in power system operational applications. They are essentially sensitivity factors that measure the incremental change in branch flow due to a change in power transferring between a slack bus and a non-slack bus. Thus, given a branch $k$, the loads that have significant impacts on that branch should be monitored. It would be unusual if the changes in all the loads that are critical to branch $k$ contribute to decreasing the flow on branch $k$. Therefore, based on this observation, a malicious load deviation index is proposed to detect potential FDI cyber-attacks. $MLDI_k$ is defined in (21),

$$MLDI_k = sgn(P_{k-})\frac{\sum_{n \in NL(k)} Indictr_{n,k}}{NL_k} \quad (21)$$

where,

$$Indictr_{n,k} = \begin{cases} -sgn(PTDF_{n,k}), & if\ \frac{d_{n0,M}-d_{n-}}{d_{n-}} \leq -5\% \\ 0, & if\ -5\% < \frac{d_{n0,M}-d_{n-}}{d_{n-}} < 5\% \\ sgn(PTDF_{n,k}), & if\ \frac{d_{n0,M}-d_{n-}}{d_{n-}} \geq 5\% \end{cases} \quad (22)$$

and

$$NL_k = size(NL(k)) \quad (23)$$

where $NL(k)$ denotes the load buses that are critical to branch $k$. If the absolute value of $PTDF_{n,k}$ is greater than or equal to 1%, then, the associated load bus $n$ is defined to be critical to branch $k$. The indicator, $Indictr_{n,k}$, defined in (22), measures the impact of load change at bus $n$ on the flow change on branch $k$; it is set to zero if the impact is insignificant.

Though theoretically $MLDI_k$ is in the range of [-1, 1], it should be close to zero if loads fluctuate randomly. A positive value indicates that the load change may decrease the flow on branch $k$. A very high positive value may imply the load fluctuation is abnormal and the probability of branch $k$ being targeted by an FDI attack is high.

Metric $MLDI_k$ only considers the number of load buses that are critical to an individual branch, but it fails to take load magnitude and PTDF values into account. To consider those two factors, an enhanced malicious load deviation index (EMLDI) is proposed in this work. $EMLDI_k$ is defined in (24),

$$EMLDI_k = sgn(P_{k-})\sum_{n \in NL(k)}(w_{n,k} Indictr_{n,k}) \quad (24)$$

where $w_{n,k}$ denotes the influential factor for branch $k$ due to the change in the load at bus $n$ and it can be calculated by (25).

$$w_{n,k} = \frac{|(d_{n0,M}-d_{n-})PTDF_{n,k}|}{\sum_{n \in NL(k)}|(d_{n0,M}-d_{n-})PTDF_{n,k}|} \quad (25)$$

$EMLDI_k$ shares the same range and indication with $MLDI_k$. However, load magnitude and PTDF values are not considered in $MLDI_k$ but are captured by $EMLDI_k$. Thus, given a specific potential target branch $k$, $EMLDI_k$ may be a better indicator to determine whether there is an attack targeting that branch. Though the alert level criteria for $MLDI_k$ and $EMLDI_k$ defined in Table II are the same, $EMLDI_k$ should be used for determining the alert level since it captures more factors than $MLDI_k$. $ALE_k$ denotes the alert level associated with $EMLDI_k$.

TABLE II ALERT LEVEL CRITERIA BASED ON $EMLDI_k$

| Alert level $ALE_k$ | $MLDI_k$ or $EMLDI_k$ |
|---|---|
| *Danger* | >50% |
| *Warning* | >35% |
| *Monitor* | >20% |
| *Normal* | <20% |

### B. Two-stage FDID Approach

Metrics MLDI and BORI presented in Section IV.A are used to detect potential FDI cyber-attack on a specific branch rather than to monitor the system as a whole. Thus, a systematic two-stage FDID approach, consisting of an FDI attack awareness stage and a target branch identification stage, is proposed in this paper to detect FDI cyber-attack. The first stage is to determine whether the system is under FDI cyber-attack and the second stage would identify the target branch.

**Stage 1: FDI Attack Awareness**

MLDI and BORI are proposed to detect whether an FDI cyber-attack is launched for a specific branch. Since system operators have limited information regarding which branch the attacker would target, it is necessary to calculate the metrics for all branches. However, given that a practical power system typically has a large number of branches, even random load fluctuations may cause large values of $MLDI_k$, $EMLDI_k$, and $BORI_k$ for a few branches, which may mislead system operators to believe that the load fluctuation is abnormal and the system is under attack. Therefore, a system-wide malicious load deviation index (SMLDI) is proposed to resolve this issue. SMLDI is defined in the equation shown below,

$$SMLDI = \frac{\sum_{k \in KA} MLDI_k}{size(KA)} \quad (26)$$

where $KA$ is a set of ten branches that have top ten $MLDI_k$ values. If the number of load buses that have significant effects on branch $k$ is too small, then the associated $MLDI_k$ are not used for malicious load deviation recognition and branch $k$ will not be included in the set $KA$. In this work, branches that have less than five critical load buses will not be considered as a candidate element of set $KA$.

TABLE III ALERT LEVEL CRITERIA BASED ON $SMLDI$

| Alert level | $SMLDI$ |
|---|---|
| *Danger* | >50% |
| *Warning* | >35% |
| *Monitor* | >20% |
| *Normal* | <20% |

In this stage, SMLDI is used as the metric to determine whether the system is under attack. Similar to the alert level designed for a target branch, a system-wide FDI cyber-attack alert level is defined in Table III. A system would be considered to be FDI cyber-attack free if the associated alert level is marked as *Normal* or *Monitor* in the first stage. Only the cases that have either *Warning* or *Danger* alert flags will be sent to the second stage for target branch identification.

**Stage 2: Target Branch Identification**

It is vital to determine whether the system is under malicious FDI cyber-attack in Stage 1. It is also very important to identify the branch that the attacker targets so that operators can take immediate actions to handle the detected attack.

$EMLDI_k$ detects the attack by monitoring suspicious load deviations while $BORI_k$ detects FDI attacks from the viewpoint of potential flow violations. The alert level $ALE_k$ associated with $EMLDI_k$ and the alert level $ALB_k$ associated with $BORI_k$ can be combined into a single comprehensive attack alert level, which is defined in Table IV. This combined alert level, denoted by $ALC_k$, is used to identify the target branch.

Though the proposed alert system can provide a qualitative analysis, it is also very important to analyze the FDI cyber-attack quantitatively. Thus, a comprehensive attack index (CAI) that considers both load deviation patterns and potential branch overloads is proposed in this work. $CAI_k$ is defined in (27). The branches that have the largest $CAI_k$ are considered to be the most suspicious target branches. Moreover, the $CAI_k$ rank indicates the possibility of branch $k$ being targeted.

$$CAI_k = EMLDI_k \, BORI_k \quad (27)$$

Therefore, both the proposed comprehensive attack index $CAI_k$ and the proposed comprehensive alert level $ALC_k$ will be used to identify the target branch in Stage 2. The branches that are either marked as *Danger* or have a $CAI_k$ ranking in the top three are considered as the most suspicious target branches.

TABLE IV DETERMINATION OF THE COMPREHENSIVE ALERT LEVEL $ALC_k$

| $ALC_k$ | $ALE_k$ | | | |
|---|---|---|---|---|
| $ALB_k$ | Normal | Monitor | Warning | Danger |
| Normal | *Normal* | *Monitor* | *Monitor* | *Warning* |
| Monitor | *Monitor* | *Monitor* | *Warning* | *Warning* |
| Warning | *Monitor* | *Warning* | *Warning* | *Danger* |
| Danger | *Warning* | *Warning* | *Danger* | *Danger* |

### C. Enhanced EMS with FDID

Since the proposed FDI cyber-attack detection is independent with existing EMS applications, it can be considered as a separate module that processes measurements in the first place before state estimation runs. The added FDID module will neither change the current EMS structure nor each single EMS application. The flowchart shown in Fig. 4 illustrates the proposed FDID method and how it integrates with current power system real-time operations. If no FDI cyber-attack events are identified, EMS will execute ordinarily; otherwise, special protection schemes will be activated to handle the identified cyber-attack events. For instance, a higher value of the initial line flow and/or a lower value of the line capacity can be used for the attacked target line to offset the impact of the FDI cyber-attack since tampered measurements may not be identified and corrected instantly. This paper focuses on the real-time situation awareness for system operators, and the special protection scheme under FDI cyber-attack that can be considered as a third stage to the proposed two-stage FDID approach is out of the scope of this work.

## V. CASE STUDIES

The IEEE 118-bus test system is used in this paper to investigate the proposed FDI cyber-attack model and examine the proposed two-stage FDID approach. This system has 118 buses, 186 branches, and 19 online units. Out of 118 buses, 99 buses are load buses. The initial total load is 4,242 MW. The detailed result analysis on the IEEE 118-bus test system are presented in Section V.A and Section V.B. In addition, robustness analysis is presented in Section V.C to demonstrate the developed FDID approach with the proposed metrics on the same IEEE 118-bus test system under different network configurations and two other different test systems: IEEE RTS96 system and the Polish system.

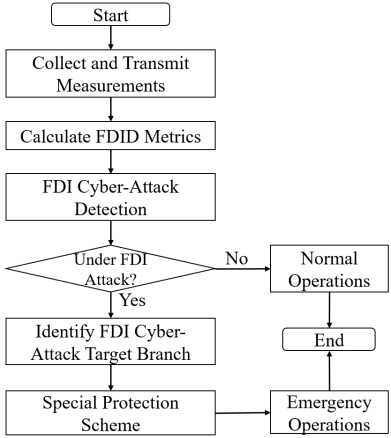

Fig. 4. Flowchart of the proposed FDID method.

### A. FDI Results

To study the effectiveness of the proposed FDI cyber-attack model, numerical simulations are conducted with different scenarios including constant load scenarios and random load fluctuation scenarios in the first dispatch interval. The effects of different load shift factors and $l_1$-norm constraint limits on the physical consequences of an FDI cyber-attack are analyzed. After all possible branches are studied, branch 111 and branch 118 are selected as the target branch to illustrate the proposed work since these two branches correspond to the worst two cyber-attacks that cause the most severe overloads.

With the assumption that load profile does not change, the resulted power flows on branch 111 and branch 118 at $t = 0$ with different attack settings are presented in Fig. 5 and Fig. 6 respectively. The blue curve with diamond markers in Fig. 5 corresponds to the FDI results with a load shift factor of 5% and it becomes flat very quickly. The reason is that the load shift constraint becomes binding when $N_1$ increases to 6 and further relaxing the $l_1$-norm constraint will not affect the results. Since the FDI model used in this work is a fast heuristic rather than an exact approach, the flows shown in Fig. 5 and Fig. 6 do not strictly increase with the load shift factor and the $l_1$-norm constraint limit. However, with more flexible conditions, attacker can typically cause more severe flow violations. Note that the attacker is assumed to have access to all load measurements to simulate the worst scenario case. As a result, most load measurements, if not all, are altered to maximize the impact of the FDI cyber-attack. For instance, with $N_1$ being 5 and load shift factor being 10%, out of 99 nodal load measurements, 94 nodal load measurements are tampered to achieve the maximum overload on the target branch 118.

In reality, loads fluctuate all the time. Thus, it is very important to analyze the effects of random load fluctuations on FDI cyber-attacks. It is assumed that load fluctuation follows the normal distribution with a mean of $\mu$ (a percentage) and a standard deviation of $\sigma$ (a percentage), which is denoted by $N(\mu, \sigma)$. The process of generating a load fluctuation vector following $N(\mu, \sigma)$ is presented below:

1) generate a vector $v$ that follows standard normal distribution.
2) apply a cutoff value 1.96 to this vector $v$.
3) adjust $v$ with equation: $v = v\sigma + \mu$.
4) create a load fluctuation vector: $\Delta d_n = d_{n-}v_n, \forall n$.

Note that since loads do not fluctuate significantly in a short-term, the second step would ensure the random load fluctuation does not have a long tail distribution. The cutoff value 1.96 corresponds to a confidence interval of 95%.

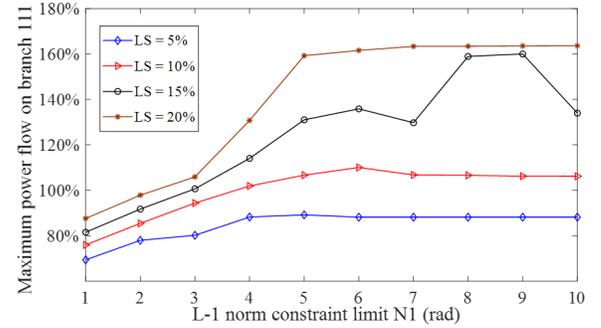

Fig. 5. Maximum power flow on branch 111 with constant load under FDI cyber-attacks with various load shift factors and $l_1$-norm constraint limits.

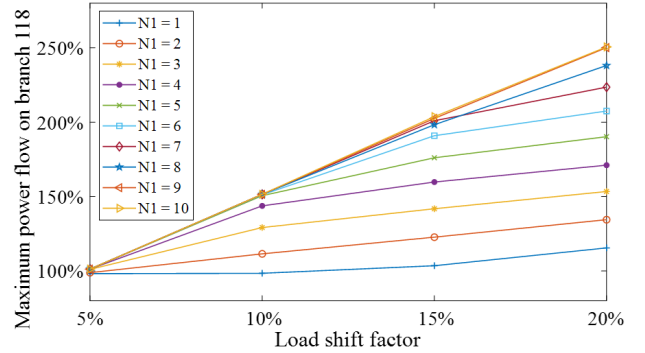

Fig. 6. Maximum power flow on branch 118 with constant load under FDI cyber-attacks with various load shift factors and $l_1$-norm constraint limits.

For the FDI simulations conducted in this section, the random load fluctuations follow the normal distribution of $N(0, 3\%)$. For each attack simulated, the load profile will be updated with a different randomly generated load fluctuation vector. The results of FDI cyber-attacks with load fluctuation are presented in Fig. 7 and Fig. 8. Fig. 7 shows the results of attacks targeting branch 111 while Fig. 8 shows the results of attacks targeting branch 118. The curves in Fig. 7 and Fig. 8 look very similar to the corresponding curves in Fig. 5 and Fig. 6 respectively. This indicates that the impact of random load fluctuation on FDI cyber-attack is limited. Fig. 7 and Fig. 8 show that an FDI cyber-attack can still result in a flow violation on the target branch even with random load fluctuations.

### B. FDID Results
*Stage 1: FDI Attack Awareness*

The proposed FDID strategy consists of two stages. Stage 1 determines whether the system is under an FDI cyber-attack by analyzing the load profile change pattern. It is important to detect the attack and it is also vital to bypass normal random load fluctuations. The goal of Stage 1 is to have a low probability of false alarm and a low probability of false dismissal. Two sets of system scenarios with different load deviation vectors, including FDI malicious load deviation vectors and random load fluctuation vectors, are tested in this stage.

The load deviation vector denotes the difference between the loads at the beginning of the second dispatch interval ($t = 0^+$) and the loads at the beginning of the first dispatch interval. The malicious load deviation vectors are obtained from the 160 different attacks performed in Section IV.A. The normal load fluctuation vectors are created with four different normal distributions: $N(0, 3\%)$, $N(0, 5\%)$, $N(-1\%, 3\%)$, and $N(1\%, 3\%)$. Twenty independent vectors are generated for each normal distribution and thus, the second set of system scenarios correspond to 80 load fluctuation vectors.

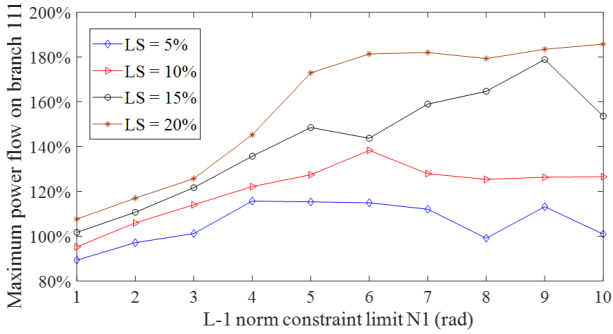

Fig. 7. Maximum power flow on branch 111 with random load fluctuation under FDI cyber-attacks.

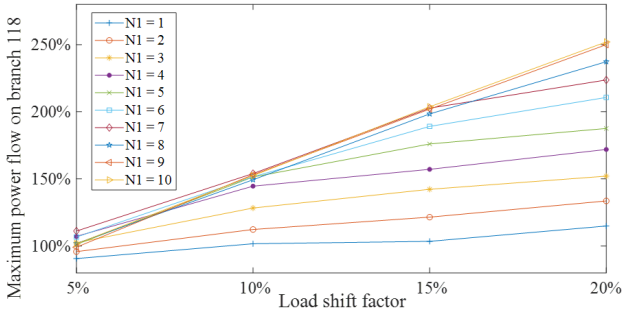

Fig. 8. Maximum power flow on branch 118 with random load fluctuation under FDI cyber-attacks.

The SMLDI values for those 240 scenarios are calculated in Stage 1. The results for the 160 FDI malicious load deviation vectors and the 80 random load fluctuation vectors are presented in Table V and Table VI respectively. The SMLDI values for random load fluctuations are very small and the averages are close to zero. However, the SMLDI values with attack are much bigger and the average values are around 70%. This indicates that FDI cyber-attacks are successfully detected with the proposed metric SMLDI and random load fluctuations can successfully bypass the proposed FDID approach.

Fig. 9 shows a scatter plot of the SMLDI values for the random load fluctuations and FDI cyber-attacks. The blue squares correspond to the random load fluctuations while the red triangles correspond to the FDI cyber-attacks. As shown in Fig. 9, the SMLDI values for random load fluctuations are all below 35% and the associated alert levels are either *Normal* or *Monitor*. Moreover, the alert levels for most random vectors are *Normal*. As for the FDI cyber-attacks, the associated SMLDI values are all above the *Warning* tolerance and the alert levels for most attacks are *Danger*. This demonstrates the proposed metric SMLDI can efficiently detect FDI cyber-attack and would not mistakenly identify a random load fluctuation as an FDI cyber-attack. In other words, the results presented in Fig. 9 demonstrate the proposed FDID scheme has a low false alarm rate as well as a low false dismissal rate.

The first 80 system scenarios in Fig. 9 correspond to random load fluctuations with four different normal distributions. They are listed in the order of $N(0, 3\%)$, $N(0, 5\%)$, $N(-1\%, 3\%)$, and $N(1\%, 3\%)$. Each normal distribution has 20 scenarios. By comparing the random load fluctuations generated with different normal distributions, it is observed that the mean of load fluctuation does not significantly affect the metric while higher standard deviations may result in higher SMLDI values. This is consistent with the statistics presented in Table VI. This implies that the false alarm rate for the proposed approach would increase as the magnitude of load fluctuation increases. It is worth noting that loads typically do not deviate substantially in a short time frame.

TABLE V SMLDI VALUES FOR FDI CYBER-ATTACKS

|  | Attack on branch 118 | | Attack on branch 111 | |
|---|---|---|---|---|
|  | Constant load | $N(0, 3\%)$ | Constant load | $N(0, 3\%)$ |
| max | 97.8% | 97.8% | 97.8% | 97.5% |
| min | 48.8% | 39.9% | 35.7% | 38.7% |
| median | 62.9% | 68.5% | 62.9% | 63.8% |
| average | 72.1% | 74.5% | 68.6% | 70.7% |
| std | 16.8% | 19.0% | 19.4% | 20.3% |

TABLE VI SMLDI VALUES FOR RANDOM LOAD FLUCTUATIONS

|  | Normal load fluctuation and no FDI attack | | | |
|---|---|---|---|---|
|  | $N(0, 3\%)$ | $N(0, 5\%)$ | $N(-1\%, 3\%)$ | $N(1\%, 3\%)$ |
| max | 23.1% | 28.0% | 23.5% | 20.9% |
| min | 3.2% | 13.8% | 5.5% | 7.5% |
| median | 11.8% | 22.8% | 12.0% | 12.5% |
| average | 12.2% | 21.7% | 12.4% | 13.2% |
| std | 4.4% | 3.7% | 4.7% | 3.7% |

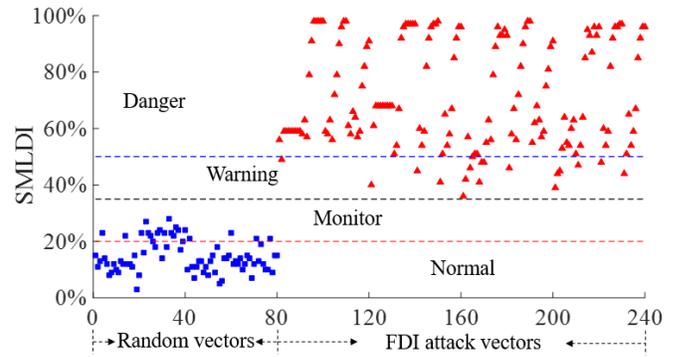

Fig. 9. SMLDI values for random load fluctuations and FDI cyber-attacks.

Fig. 10 illustrates the SMLDI values that are associated with various FDI attacks targeting branch 118 with random load fluctuations that follow $N(0, 3\%)$. The red dotted straight line is the boundary between the alert levels *Monitor* and *Warning*. Those SMLDI values are well above the *Warning* alert tolerance of 35%, especially for the cases that have more flexible attack constraints. It is very straightforward and efficient to identify whether the system is under malicious FDI cyber-attack with the proposed metric SMLDI.

The results shown in Fig. 9 and Fig. 10 correspond to the metric SMLDI defined in (26), where the set *KA* consists of 10 branches that have top ten $MLDI_k$ values. In order to further justify the proposed metric, sensitivity of the metric performance to the number of branches in set *KA* for calculating SMLDI is conducted; Fig. 11 shows the SMLDI values with 8 branches while Fig. 12 shows the SMLDI values with 12 branches. It is observed that the results shown in Fig. 11 and Fig. 12 closely match the results shown in Fig. 9 and they all show there is a clear line cut between FDI cyber-attack scenarios and normal random load fluctuation scenarios. This indicates that the proposed metric performance is not very sensitive to the number of branches for calculating SMLDI. To be consistent, for the rest of this paper, *KA* is defined as a set of 10 branches that have the highest ten $MLDI_k$ values.

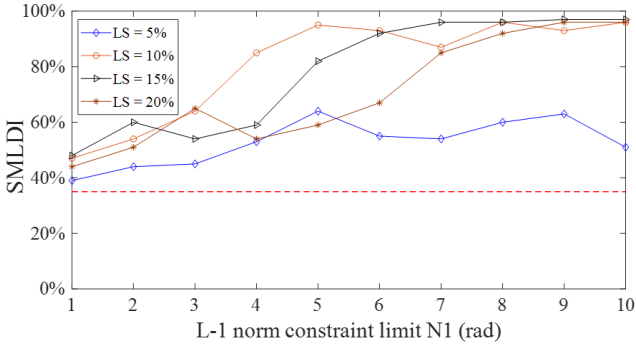

Fig. 10. SMLDI of FDI attacks targeting branch 118 with a random load fluctuation of *N*(0, 3%).

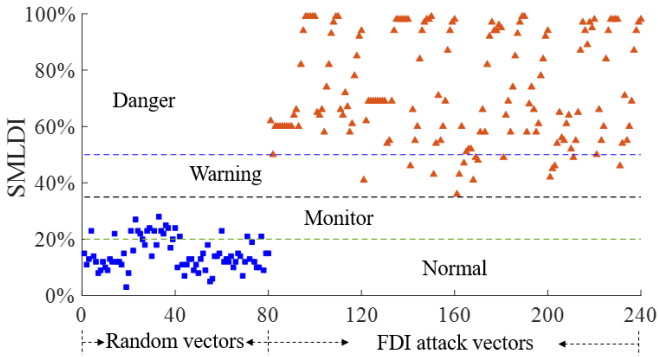

Fig. 11. SMLDI values for random load fluctuations and FDI cyber-attacks when set *KA* consists of 8 branches that have top 8 $MLDI_k$ values.

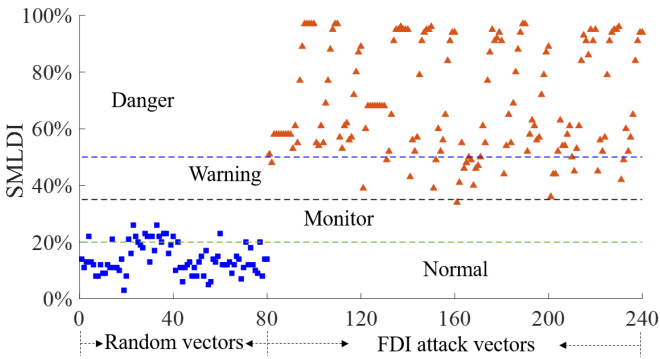

Fig. 12. SMLDI values for random load fluctuations and FDI cyber-attacks when set *KA* consists of 12 branches that have top 12 $MLDI_k$ values.

*Stage 2: Target Branch Identification*

In the second stage, only the cases that are identified to be under FDI cyber-attack will be examined. Thus, only those 160 FDI attack scenarios identified in the first stage are sent to the target branch identification routine.

Table VII shows the results of target branch identification against the FDI attacks on branch 111 with no random load fluctuations and a load shift factor of 10% in the attack model. The metric $CAI_k$ for branch 111 ranks first for nine scenarios out of the ten scenarios and ranks second for the remaining one scenario. There are eight scenarios for which a branch marked as *Danger* exists; and branch 111 is the only one that is marked as *Danger* for those eight scenarios. Therefore, both the proposed comprehensive FDI attack index and the proposed comprehensive alert level successfully indicate branch 111 is the most suspicious target.

Table VIII shows the results of target branch identification against the FDI attacks on branch 111 with a random load fluctuation that follows *N*(0, 3%) and a load shift factor of 10% in the attack model. The simulations corresponding to Table VII do not involve random load fluctuations while the simulations for Table VIII do. Thus, the results shown in Table VIII are more realistic. However, the conclusions drawn from Table VIII are consistent with Table VII. This indicates that the proposed strategy is also effective even when load fluctuation is considered.

TABLE VII TARGET BRANCH IDENTIFICATION RESULTS FOR FDI ATTACKS ON BRANCH 111 WITH NO RANDOM LOAD FLUCTUATIONS AND A LOAD SHIFT FACTOR OF 10% IN THE ATTACK MODEL

|  | $CAI_{111}$ | Rank of $CAI_{111}$ | $ALC_{111}$ | Number of lines marked *Danger* |
|---|---|---|---|---|
| $N_1 = 1$ | 0.46 | 2 | *Monitor* | 0 |
| $N_1 = 2$ | 0.66 | 1 | *Warning* | 0 |
| $N_1 = 3$ | 0.90 | 1 | *Danger* | 1 |
| $N_1 = 4$ | 1.13 | 1 | *Danger* | 1 |
| $N_1 = 5$ | 1.26 | 1 | *Danger* | 1 |
| $N_1 = 6$ | 1.30 | 1 | *Danger* | 1 |
| $N_1 = 7$ | 1.27 | 1 | *Danger* | 1 |
| $N_1 = 8$ | 1.27 | 1 | *Danger* | 1 |
| $N_1 = 9$ | 1.26 | 1 | *Danger* | 1 |
| $N_1 = 10$ | 1.26 | 1 | *Danger* | 1 |

TABLE VIII. TARGET BRANCH IDENTIFICATION RESULTS FOR FDI ATTACKS ON BRANCH 111 WITH *N*(0, 3%) RANDOM LOAD FLUCTUATION IN THE FIRST INTERVAL AND A LOAD SHIFT FACTOR OF 10% IN THE ATTACK MODEL

|  | $CAI_{111}$ | Rank of $CAI_{111}$ | $ALC_{111}$ | Number of lines marked *Danger* |
|---|---|---|---|---|
| $N_1 = 1$ | 0.45 | 2 | *Monitor* | 0 |
| $N_1 = 2$ | 0.68 | 1 | *Warning* | 0 |
| $N_1 = 3$ | 0.91 | 1 | *Danger* | 1 |
| $N_1 = 4$ | 1.07 | 1 | *Danger* | 1 |
| $N_1 = 5$ | 1.21 | 1 | *Danger* | 1 |
| $N_1 = 6$ | 1.37 | 1 | *Danger* | 2 |
| $N_1 = 7$ | 1.19 | 1 | *Danger* | 1 |
| $N_1 = 8$ | 1.29 | 1 | *Danger* | 2 |
| $N_1 = 9$ | 1.23 | 1 | *Danger* | 1 |
| $N_1 = 10$ | 1.20 | 1 | *Danger* | 1 |

Table IX presents the FDID results on various FDI attacks. As shown in this table, the target branches are correctly identified for 96.9% or 155 out of 160 FDI cyber-attacks. The target branch is marked as *Danger* for over 90% of the FDI attacks on branch 118. The percentage of the cases that the target branch of the FDI attacks on branch 111 is marked as *Danger* is relatively low. The reason is that the overloads on branch 111 for most FDI attacks with a load shift factor of 5% are insignificant and do not reach the *Warning* alert threshold. However, the associated comprehensive FDI attack index of the target branch 111 ranks first for most cases. For all FDID tests on the 160 FDI attacks, the comprehensive attack indices of the target branch rank very high and almost all of them rank either first or second.

TABLE IX RESULTS OF FDID ON VARIOUS FDI ATTACKS

|  |  | Average $CI_k$ rank of the target branch | Percent of scenarios for which the target branch is identified | Percent of scenarios for which the target branch is marked as *Danger* | # of scenarios simulated |
|---|---|---|---|---|---|
| Attack on branch 118 | Constant load | 1.58 | 92.5% | 92.5% | 40 |
|  | $N(0, 3\%)$ | 1.55 | 100% | 92.5% | 40 |
| Attack on branch 111 | Constant load | 1.13 | 100% | 65% | 40 |
|  | $N(0, 3\%)$ | 1.33 | 95.0% | 77.5% | 40 |
| Cumulative statistics |  | 1.39 | 96.9% | 81.9% | 160 |

### C. Robustness Analysis

To validate the robustness of the proposed FDID metrics and approach and show that they can be generalized to different network configurations and different systems, this subsection conducts extensive robustness analysis studies.

*IEEE 118-Bus System under Different Configurations*

It is very common that practical power systems experience outages including planned outage for component maintenance and unplanned outage that is also referred to as contingency [26]. This indicates that the network configuration of a practical power system varies in different time periods. Thus, it is very important to examine the proposed approach under different grid configurations. In this work, three configurations denoted as CNFRTN-1, CNFRTN-2, and CNFRTN-3 are studied and they correspond to three random single outages: an outage on branch 1, an outage on branch 71, and an outage on branch 141 respectively. Normal random load fluctuations without cyber-attack and FDI cyber-attacks on the same branch 111 and branch 118 with and without noises are simulated on the 118-bus test system under those three configurations respectively. The normal load fluctuation scenarios are represented by four sets of ten independent vectors generated for four normal distributions respectively: $N(0, 3\%)$, $N(0, 5\%)$, $N(-1\%, 3\%)$, and $N(1\%, 3\%)$. The 32 FDI attack vectors corresponds to two target branches under various attack conditions (with $N_1$ being 5 and 10 respectively, and load shift factor being 5%, 10%, 15% and 20% respectively).

The resulted power flows on the target branch 118 under the base case configuration and three variations are illustrated in Fig. 13. From this figure, we can conclude that the network configuration change does not significantly affect the FDI cyber-attack results on maximizing the target branch flow. The FDID results under different system network configurations are shown in Table X. It is observed that the proposed FDID method can effectively detect the malicious FDI cyber-attack scenarios and will not mistakenly identify normal random load fluctuation scenarios as FDI cyber-attacks; moreover, the proposed FDID approach achieves very similar results under different configurations.

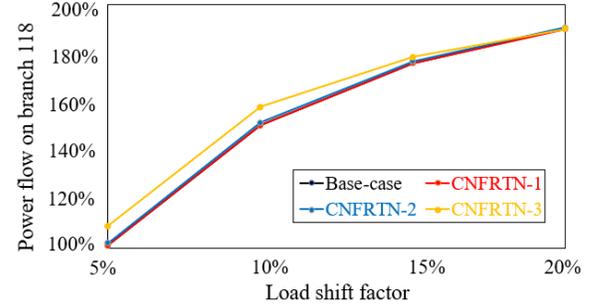

Fig. 13. Maximum power flow on branch 118 with constant load under FDI cyber-attacks against different network configurations.

TABLE X RESULTS OF FDID ON DIFFERENT SYSTEM CONFIGURATIONS

|  | Stage 1: Random load fluctuations scenarios |  |  |  |  | Stage 1: FDI cyber-attack scenarios |  |  |  |  | Stage 2 |  |
|---|---|---|---|---|---|---|---|---|---|---|---|---|
|  | # of scenarios simulated | $Pcntg_1$ | SMLDI |  |  | # of scenarios simulated | $Pcntg_2$ | SMLDI |  |  | # of scenarios simulated | $Pcntg_3$ |
|  |  |  | max | min | avg |  |  | max | min | avg |  |  |
| CNFRTN-1 | 40 | 0% | 28% | 7% | 15% | 32 | 100% | 98% | 47% | 79% | 32 | 100% |
| CNFRTN-2 | 40 | 0% | 29% | 7% | 15% | 32 | 100% | 98% | 47% | 79% | 32 | 100% |
| CNFRTN-3 | 40 | 0% | 29% | 7% | 15% | 32 | 100% | 98% | 48% | 77% | 32 | 100% |

Note that $Pcntg_1$ denotes the percentage of normal random load fluctuation scenarios that are mistakenly identified as FDI cyber-attacks; $Pcntg_2$ denotes the percentage of malicious FDI cyber-attack scenarios that are successfully detected; $Pcntg_3$ denotes the percentage of FDI cyber-attack scenarios of which the target branches are identified.

*IEEE RTS-96 System*

The IEEE RTS-96 system is a commonly studied test system that has 73 buses. Two highly loaded branches are selected as the target: branch 62 and branch 99. This work simulated a total of 40 FDI cyber-attack scenarios that correspond to the two target branches with and without consideration of load fluctuations under various attack conditions (with $N_1$ being 1 through 10 respectively, and load shift factor being 10%). The 40 normal load fluctuation scenarios are created with the same procedure for the IEEE 118-bus system under different configurations.

Fig. 14 shows branch overloads in MW due to FDI cyber-attack. *Ln62c* and *Ln99c* represent the overload on branch 62 and branch 99 respectively with the assumption that load does change while *Ln62f* and *Ln99f* show the overload when considering load fluctuation. The *Ln62c* and *Ln99c* curves are non-decreasing as the attack parameter $N_1$ increases. As demonstrated by the *Ln62f* and *Ln99f* curves, load fluctuations may only slightly affect the attack impact.

Fig. 15 presents the SMLDI values for normal load fluctuation scenarios and FDI cyber-attack scenarios on the IEEE RTS-96 73-bus test system. We can observe that the metric levels used for the IEEE 118-bus test system can also clearly distinct the FDI cyber-attacks and normal load fluctuations on this IEEE 73-bus system. All FDI cyber-attacks are detected and target branches are identified for this test system.

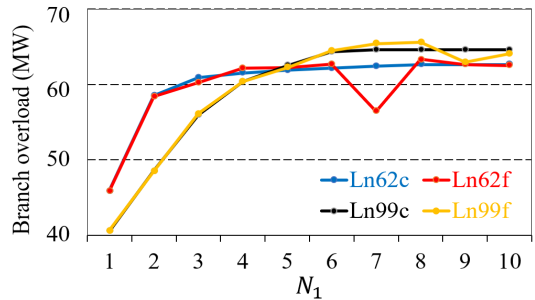

Fig. 14. Branch overload under FDI cyber-attacks on the 73-bus test system.

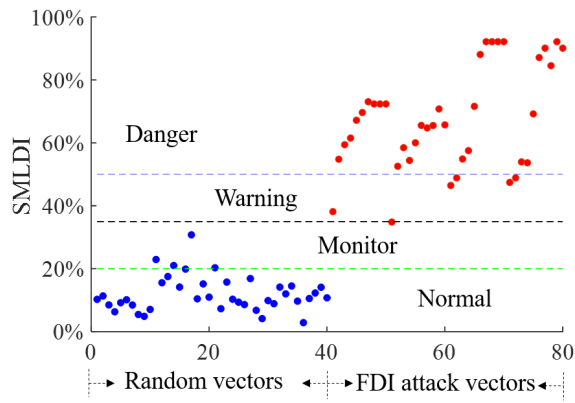

Fig. 15. SMLDI values for random load fluctuations and FDI cyber-attacks on the IEEE RTS-96 73-bus test system.

*Polish System*

Numerical simulations are conducted on the Polish system that has 2,383 buses to further demonstrate the robustness of the proposed approach. Five highly loaded branches are selected as the FDI cyber-attack target. Sixteen FDI cyber-attack scenarios are designed for each target branch; they differ in the attack parameters ($L_S$ and $N_1$) and load profiles. The 80 normal load fluctuation scenarios are created with the same procedure as Section V.B. Simulation results show the FDI cyber-attacks can cause violations on the Polish system; for example, it leads to an average of 15.2 MW branch overload when attack parameter $L_S$ is set to 10% and $N_1$ is set to 10 and load fluctuation is ignored.

Table XI presents the FDI cyber-attack detection results. The average and median of the proposed SMLDI metric values for 80 FDI cyber-attack scenarios are both over 70%, which is way above the warning alert level 35%; the minimum SMLDI metric value is 38% that will also trigger the alert system and the associated FDI cyber-attack will be sent to Stage 2 for target branch identification. In Stage 2, the target branch are corrected identified for 74 (92.5%) attack scenarios when using the comprehensive attack index only; the target for 54 (67.5%) attack scenarios are marked as *DANGER* that implies the associated branch is highly likely to be the target. Overall, the target branch identification in the second stage has a success rate of 97.5%. As for the 80 normal load fluctuation scenarios, only one single scenario with the metric SMLDI being 36% (slightly above 35%) is mistakenly identified as a potential FDI cyber-attack.

*D. Discussions*

In this case studies section, the proposed two-stage FDID strategy is demonstrated with numerical simulations on three test systems and the variations with different configurations of the 118-bus system. The proposed metrics and alert system can efficiently detect potential FDI cyber-attacks by monitoring abnormal load deviations and flow changes. Thus, we can conclude that the proposed two-stage FDID approach is effective and robust. One key factor that affects the proposed detection method is the selection of metric level thresholds. A good practice is to train enough datasets including both normal load fluctuation vectors and FDI cyber-attack vectors, calculate the metric values for both vector sets, and then a cutoff line (metric tolerance) can be determined to clearly separate the two types of scenarios. The optimal metric thresholds for different systems can be very similar, which has been demonstrated in this work; the same metric tolerances can effectively detect all FDI cyber-attacks and filter out normal load fluctuations on three very different systems: the IEEE RTS-96 73-bus system, the IEEE 118-bus system, and the Polish 2383-bus system.

TABLE XI RESULTS OF FDID ON THE POLISH SYSTEM

|  |  |  | Random load fluctuations scenarios | FDI cyber-attack scenarios |
|---|---|---|---|---|
| Stage 1 | SMLDI | max | 36% | 92% |
|  |  | min | 10% | 38% |
|  |  | median | 23% | 73% |
|  |  | average | 23% | 71% |
|  |  | std | 6% | 16% |
|  | # of scenarios simulated |  | 80 | 80 |
|  | # of scenarios identified as FDI cyber-attack |  | 1 | 80 |
| Stage 2 | # of scenarios that the target ranks top 3 of *CAI* |  | N/A | 74 (92.5%) |
|  | # of scenarios that the target is marked as *DANGER* |  |  | 54 (67.5%) |
|  | Overall success rate |  |  | 78 (97.5%) |

## VI. CONCLUSIONS

An FDI cyber-attack model is first introduced in this paper to examine the effects of FDI attacks on system reliability. Then, a two-stage systematic approach that does not rely on meter protection or historical data is proposed to detect FDI cyber-attack. The proposed FDI detection method is based on an alert system with the devised novel metrics. Two categories of metrics, MLDI and BORI, are proposed in this two-stage approach to determine whether the change in system condition is abnormal. MLDI recognizes malicious load changes while BORI identifies suspicious flow changes. Sensitivity factors like PTDFs are used to calculate the regular MLDI and the enhanced MLDI. In the first stage, the proposed system-wide MLDI is used to determine whether the system is under attack. If the system is deemed to be under attack, the second stage will execute and the proposed alert system along with the proposed comprehensive FDI attack index will be used to identify the attack target branch.

Simulation results show that FDI cyber-attack can cause physical flow violations and demonstrate the effectiveness of the proposed FDID metrics, FDI cyber-attack alert system and two-stage FDID approach. The proposed two-stage FDID approach successfully detects all FDI cyber-attacks that are simulated in this work and correctly identifies the target branch with a success rate of over 95%. In addition, random load fluctuations would not trigger the alert system. Numerical simula-tions show that almost none of random load fluctuation scenarios is mistakenly identified as malicious cyber-attacks. To conclude, normal load fluctuations will not activate the proposed FDI alert system while the proposed two-stage FDID approach can efficiently detect FDI cyber-attack and the target branch. In other words, the false alarm rate and false dismissal rate for the proposed two-stage FDID approach are low.

## VII. FUTURE WORK

Although this paper has demonstrated the performance of the proposed metric-based FDI cyber-attack alert system and the effectiveness of the developed two-stage systematic detection approach, there are still room for improvement to this work. Key future work beyond this paper is to: (i) replace the simplified DC SE based FDI model with an enhanced FDI model using full AC SE that is consistent with industrial practice and is more accurate; (ii) add AC RTCA module in the loop to simulate the entire EMS; (iii) upgrade the current two-stage FDID approach to an enhanced three-stage FDID approach by adding a third stage for identifying tampered measurements and determining the associated special protection schemes.